\documentclass[conference]{IEEEtran}

\usepackage{amssymb,amsmath,amsfonts,amsthm}
\usepackage{mathrsfs}
\usepackage{cite}
\usepackage{graphicx}

\DeclareMathOperator{\sinc}{sinc}

\newtheorem{propi}{Proposition}

\IEEEoverridecommandlockouts

\begin{document}

\title{Distributed SIR-Aware Opportunistic Access Control for D2D Underlaid Cellular Networks}


\author{Zheng Chen and Marios Kountouris  \\ 
\normalsize Department of Telecommunications \\ 
\normalsize SUPELEC, \normalsize Gif-sur-Yvette, France\\
{\normalsize E-mail:~\{zheng.chen, marios.kountouris\}@supelec.fr}}


\maketitle

\begin{abstract}
In this paper, we propose a distributed interference and channel-aware opportunistic access control technique for D2D underlaid cellular networks, in which each potential D2D link is active whenever its estimated signal-to-interference ratio (SIR) is above a predetermined threshold so as to maximize the D2D area spectral efficiency. The objective of our SIR-aware opportunistic access scheme is to provide sufficient coverage probability and to increase the aggregate rate of D2D links by harnessing interference caused by dense underlaid D2D users using an adaptive decision activation threshold. We determine the optimum D2D activation probability and threshold, building on analytical expressions for the coverage probabilities and area spectral efficiency of D2D links derived using stochastic geometry. Specifically, we provide two expressions for the optimal SIR threshold, which can be applied in a decentralized way on each D2D link, so as to maximize the D2D area spectral efficiency derived using the unconditional and conditional D2D success probability respectively. Simulation results in different network settings show the performance gains of both SIR-aware threshold scheduling methods in terms of D2D link coverage probability, area spectral efficiency, and average sum rate compared to existing channel-aware access schemes.
\end{abstract}

\section{Introduction}
Device-to-Device (D2D) communication has recently attracted significant attention and is envisaged to be a key feature of 5G wireless networks and 3GPP LTE Rel. 12. Direct D2D communication between mobile devices in proximity without passing through the macrocellular base station (BS) or core network is a promising approach for dealing with local traffic in cellular networks.
D2D communication is also expected to improve link coverage, aggregate throughput, energy consumption, and end-to-end latency, while enabling new location-based services and robust public safety communications. An overview on D2D proximity services in 3GPPP standardization activities and of the main challenges in the design of D2D-enhanced cellular standards is given in \cite{DBLP:journals/corr/LinAGR13}. 

Depending on how the spectrum is assigned for D2D communication, it can be classified into \textit{inband} and \textit{outband} D2D, where in the former the licensed cellular spectrum is reused while in the latter unlicensed spectrum is used. Inband D2D communication can be further classified into \textit{overlaid} and \textit{underlaid} D2D. In the overlaid case, D2D users are assigned with a proportion of the available cellular spectrum, which is orthogonal to that used for cellular users. In the underlaid case, D2D mobile users opportunistically access cellular spectrum occupied by cellular users. 

In spite of the potential benefits of D2D, there are significant challenges, such as interference management, self-organization, network discovery, and resource allocation, which has to be addressed \cite{DBLP:journals/corr/AsadiWM13}. In overlaid D2D, the major issue is how to optimally allocate spectrum resources to cellular and D2D users, while the majority of work on underlaid D2D networks focuses on interference mitigation and coexistence of cellular and D2D communication in the same spectrum \cite{Fodor12}. In D2D underlaid cellular networks, cellular links experience cross-tier interference from the co-channel D2D transmissions, whereas the D2D pairs experience both inter-D2D interference and cross-tier interference from cellular transmissions.

Power control and opportunistic medium access control are two effective approaches to mitigate or harness interference caused by dense D2D link deployments. These methods are related to the concept of threshold scheduling \cite{Weber07} and spatial opportunistic ALOHA \cite{5226963}, which have been studied in wireless ad hoc networks. A random network model for D2D underlaid cellular networks is analyzed using stochastic geometry and channel-aware power control algorithms are proposed in \cite{DBLP:journals/corr/abs-1305-6161}. Different opportunistic medium access control techniques for two-tier femtocell networks are investigated in \cite{6215548} and various activation schemes, based on channel, interference, and signal-to-interference ratio (SIR) knowledge, are considered. The advantage of SIR-aware access control compared to other access schemes in terms of throughput maximization is evinced. Optimal channel and interference thresholds for distributed opportunistic scheduling are determined in \cite{DBLP:journals/corr/abs-1107-1731}. Nevertheless, the optimal SIR threshold for SIR-aware opportunistic access control has not been derived.

In this paper, we consider D2D underlaid cellular networks in which an uplink cellular user intends to communicate with the BS while multiple D2D links coexist in the same spectrum. The locations of D2D transmitters are modeled by a spatial homogeneous Poisson Point Process (PPP), as a means to analytically assess the interference using stochastic geometry. We propose a distributed SIR-aware opportunistic access control technique where the main idea is that each potential D2D link decides to transmit independently based on the estimated SIR value at each potential receiver so as to maximize the D2D area spectral efficiency. We provide analytical expressions for the optimal access probability and the optimal SIR threshold for the cases where aggregate throughput is maximized based on the unconditional and the conditional coverage probability. A key contribution of this paper is to derive a simple expression for the optimal access probability and SIR threshold based on an approximation of the conditional success probability of active D2D links. Simulation results show the performance gains of optimized SIR-aware threshold scheduling in terms of D2D link coverage probability, area spectral efficiency, and average sum rate compared to existing channel-aware access schemes. Specifically, the proposed approximate SIR threshold technique is shown to significantly outperform all other access control schemes.

\section{System Model}
\begin{figure}
\centering
\includegraphics[scale=0.55]{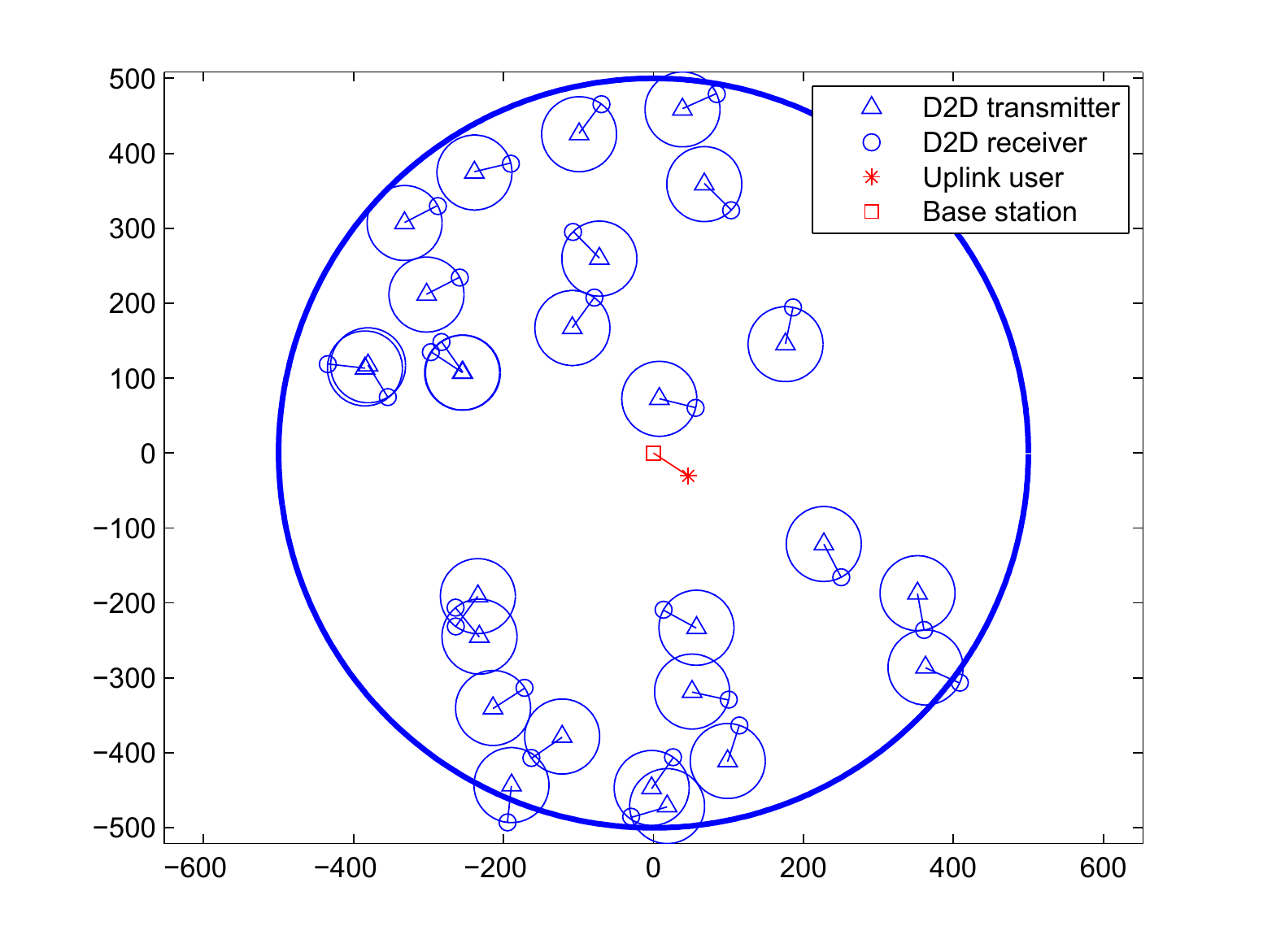}
\caption{A snapshot of a single-cell D2D underlaid cellular network with D2D link density $\lambda=3\times 10^{-5}$.}
\label{link_geometry}
\end{figure}

We consider a D2D underlaid cellular network, in which an uplink cellular user communicates with the BS while multiple direct D2D links are using the same spectrum, as is shown in Fig. \ref{link_geometry}. The coverage region of the cellular BS centered at the origin is denoted by a circular disk $\mathcal{C}$ with radius $R$. The uplink cellular user is uniformly distributed in the coverage region, while the locations of D2D transmitters are distributed in the two-dimensional Euclidean plane $\mathbb{R}^2$ according to a homogeneous spatial PPP $\Phi$ with density $\lambda$. The associated D2D receiver is distributed at random isotropic directions around its respective transmitter and at a fixed distance. Denote by $N$ the number of D2D pairs in $\mathcal{C}$, which from the assumptions above is a Poisson random variable with mean $\mathbb{E}$ $[N] = \lambda\pi R^{2}$. 

Suppose that at the beginning of each time-slot D2D transmitter-receiver pairs are able to sample the potential interfering links, thus the estimated SIR is obtained for each D2D link \cite{flashlinq}. Based on this knowledge we apply the selection of high-quality D2D links by only allowing those with relatively high estimated SIR to access spectrum.


In this paper, we consider only inter-cell interference ignoring out-of-cell interference from cellular users in other cells. Since D2D links and the uplink user share the same spectrum, the interference at a D2D receiver comes from both the uplink cellular user transmission and the other D2D transmitters.

Consider an arbitrary D2D pair with index $k\in [1,\cdots,N]$, the signal-to-interference-plus-noise ratio (SINR) at the D2D receiver is given by
\begin{equation}
 \text{SINR}_{k}(N,\textbf{p})=\frac{|{h_{k,k}}| ^{2}d_{k,k}^{-\alpha}p_{k}}{|h_{k,0}|^{2}d_{k,0}^{-\alpha}p_{c}+ \sum_{l=1,l\neq{k}}^{N} |h_{k,l}|^{2}d_{k,l}^{-\alpha}p_{l}+\sigma^{2}},
\end{equation}
where $\textbf{p}=[p_{1},\ldots, p_{N}]_{T}$ is the D2D transmitter power vector with $p_{k}$ being the power of the $k$-th D2D transmitter, $p_{k}=p_{d}$ which is a constant power value when the $k$-th transmitter is allowed to be active, whereas $p_{k} = 0$ when forbidden to transmit. The subscript 0 is used for uplink cellular transmitter with $p_{c}$ as constant transmit power, $h_{k,0}$ and $h_{k,l}$ denote small-scale amplitude fading from the uplink user to D2D receiver $k$ and from D2D transmitter $l$ to receiver $k$, which are independently distributed as $\mathcal{CN}(0,1)$, $d_{k,l}$ denotes the distance between D2D transmitter $l$ and receiver $k$. We consider a distance-dependent pathloss attenuation, which follows a standard power law, i.e. $d_{k,l}^{-\alpha}$ where $\alpha > 2$ is the pathloss exponent, $\sigma ^2$ is the complex additive noise power. 

In the remainder, we assume that the background thermal noise is negligible as compared to the self interference and is hence ignored. This is justified in the current wireless networks, which are typically interference limited \cite{BouPanJ2009}. Background noise can be included in the subsequent analytical framework with little extra work.


\section{Performance Metrics}
We present here three main performace measures we will use in order to evaluate the D2D underlaid cellular network performance using distributed opportunistic access control, namely the D2D link coverage probability, the area spectral efficiency (ASE), and the average sum rate of all D2D links.
 
\subsection{D2D Link Coverage Probability}
Consider an arbitrary communication D2D pair $k$ and assume that the D2D receiver is located at the origin. In the case where all D2D links are active, for a prescribed SIR target $\beta$, and using tools from stochastic geometry, the coverage probability of typical D2D link is given by \cite{DBLP:journals/corr/abs-1305-6161}
\begin{eqnarray}
 P^{D}_{\text{cov}}(\beta) &=& \mathbb{P}(\text{SIR}_{k}>\beta)  \nonumber\\
& =& \exp\left(-\frac{\pi\lambda d_{k,k}^{2}\beta^{\frac{2}{\alpha} }}{\sinc(\frac{2}{\alpha})}\right)\mathbb{E}\left[\frac{1}{1+ \beta \frac{p_{c}}{p_{d}} \left(\frac{d_{k,k}}{d_{k,0}}\right)^{\alpha}}\right].\nonumber\\
\label{coverage_probability}
\end{eqnarray}
Using $\mathbb{E}\left[\frac{1}{1+\frac{\kappa}{d_{k,0}^\alpha}}\right]\simeq \frac{1}{1+\frac{\kappa^{2/\alpha}}{\mathbb{E}\left[d_{k,0}\right]^2}}$ and the first moment approximation of $d_{k,0}$ as $\mathbb{E}\left[d_{k,0}\right]=\frac{128R}{45\pi}$ \cite{Moltchanov:2012:SPD:2189419.2189534} with pdf
\begin{eqnarray*}
f_{d_{k,0}} = \frac{2r}{R^{2}}\left(\frac{2}{\pi} \cos^{-1} \left(\frac{r}{2R}\right)-\frac{r}{\pi R} \sqrt{1-\frac{r^{2}}{4R^{2}}}\right), 0\leq r\leq 2R,
\end{eqnarray*}
we have
\begin{eqnarray}
 P^{D}_{\text{cov}}(\beta)
& \simeq & \exp\left(-\frac{\pi\lambda d_{k,k}^{2}\beta^{\frac{2}{\alpha} }}{\sinc(\frac{2}{\alpha})}\right)\frac{1}{1+ \left(\beta \frac{p_{c}}{p_{d}} \right)^{\frac{2}{\alpha}} \frac{d_{k,k}^{2}}{\left(128R/(45\pi) \right)^{2}}} \nonumber\\
& = &\exp\left(-\frac{\pi\lambda d_{k,k}^{2}\beta^{\frac{2}{\alpha} }}{\sinc(\frac{2}{\alpha})}\right)\frac{1}{1+K(\alpha) \beta^{\frac{2}{\alpha}}},
\label{coverage_proba}
\end{eqnarray}
where $K(\alpha)=\left(\frac{p_{c}}{p_{d}}\right)^{\frac{2}{\alpha}}\frac{d_{k,k}^{2}}{\left(128R/(45\pi)\right)^{2}}$.
As shown in \cite{DBLP:journals/corr/abs-1305-6161}, the above expression provides a very precise approximation of the D2D link coverage probability, especially when $\alpha = 4$. 

\subsection{Area Spectral Efficiency}
The area spectral efficiency (ASE), often referred to as network throughput, is a measure of spatial reuse and gives the spectral efficiency (maximum average data rate per Hz) per unit area. For the D2D underlaid cellular network considered here, the ASE can be written as  
\begin{equation}
\mathcal{T}(\beta)=\lambda \mathbb{P}(\text{SIR}_{k}>\beta)\log_{2}(1+\beta).
\end{equation}

\subsection{Average Sum Rate}
Assuming Gaussian codebooks and capacity-achieving adaptive modulation/coding, and treating interference as noise, the average achievable sum rate of D2D links is given by
\begin{eqnarray}
\mathcal{R} &=& \mathbb{E} \left[ \sum_{k=1}^{N} \log_{2} (1+\text{SIR}_{k}) \right] \nonumber \\
& = & \lambda \pi R^{2} \mathbb{E} \left [ \log_{2} (1+\text{SIR}_{k}) \right ] \nonumber \\
& = & \frac{\lambda \pi R^{2}}{\ln2} \int_{0}^{\infty} \frac{\mathbb{P}(\text{SIR}_{k}>x)}{1+x}\text{d}x.
\label{sumrate}
\end{eqnarray}
%

\section{Distributed SIR-Aware Opportunistic Access Control}
In this section we introduce a distributed opportunistic access control algorithm based on the estimated SIR value at each potential D2D receiver, where a D2D link can be activated if its estimated SIR is above the decision threshold. We propose two theoretical approaches to calculate the optimal SIR threshold that maximizes the ASE of D2D links, namely unconditional and conditional optimal thresholds, depending on how the D2D success probability is calculated. We choose the ASE rather than D2D sum rate as the performance metric to be maximized since the integral in \eqref{sumrate} adds difficulty in having closed-form solution for the optimization problem.

Denote by $G$ the nonnegative SIR threshold, a D2D pair $k$ is active, i.e. the D2D user transmits with constant power $p_d$, when the estimated link quality satisfies $\text{SIR}_{k}>G$. The estimated SIR can be obtained by assuming all potential D2D transmitters active and calculating the SIR of received signal at each D2D receiver. The set of selected D2D transmitters to be active is hence given by $\Phi_{a}=\left\{k \in (1, \ldots, N) : \text{SIR}_{k}>G\right\}$. The average access (activation) probability of a D2D transmitter, denoted by $P_{s}$, is the same as the D2D link coverage probability with $G$ as target SIR. So we have
\begin{equation}
P_{s}=\mathbb{P}(\text{SIR}_{k}>G)=\exp\left(-\frac{\pi\lambda d_{k,k}^{2} G^{\frac{2}{\alpha} }}{\sinc(\frac{2}{\alpha})}\right)\frac{1}{1+K(\alpha) G^{\frac{2}{\alpha}}}.
\label{Ps}
\end{equation}
Note that $P_{s}$ is a mean value by averaging over the fading statistics and all realizations of PPP $\Phi$. For a specific PPP realization or conditioned on $\Phi$, each D2D link experiences different SIR and thus has different access probability depending on its location and surroundings, i.e. the locations of its interferers for this realization. In other words, when there are many interferers in the vicinity of this D2D link, this link has less access probability than one in an area isolated from close interferers due to the fact that it has potentially low SIR. So for each realization of $\Phi$, we have a dependent thinning of the homogeneous PPP $\Phi$, and the set of active D2D transmitters $\Phi_{a}$ forms an inhomogeneous PPP with average density parameter $\widetilde{\lambda}=\lambda P_{s}$. For tractability and in order to derive neat expressions for the SIR threshold calculation, in the following part we work with the assumption that $\Phi_{a}$ forms a homogeneous PPP with density $\widetilde{\lambda}$.

The ASE of the D2D underlaid cellular network is given by
\begin{equation}
\mathcal{T}(\beta)= \widetilde{\lambda}\log_{2}(1+\beta) \mathbb{P}(\widetilde{\text{SIR}}_{k}>\beta|\text{SIR}_{k}>G),
\label{thoughput_definition}
\end{equation}
where $\widetilde{\text{SIR}}_{k}$ denotes the SIR of active D2D link $k \in \Phi_{a}$ and the coverage probability of a typical active D2D link is a conditional probability given that the D2D pair $k$ could communicate, i.e. its estimated $\text{SIR}_{k}$ exceeds the threshold $G$.



In the following sections, we focus on how to derive the optimal SIR threshold $G$ to maximize the ASE of the D2D underlaid network. We show that calculating $G$ without considering the conditional coverage probability in the resulting ASE, as it is the case in previous work in the literature, does not fully capture the dependence between distributed threshold scheduling at each D2D user and the resulting interference field. For that, the main contribution of this paper is to show the effectiveness of deriving the optimal $G$ based on the conditional coverage probability. 

\subsection{SIR-Aware Access Control with Unconditional Optimal SIR Threshold}
Due to the difficulty of calculating the conditional probability $\mathbb{P}(\widetilde{\text{SIR}}_{k}>\beta|\text{SIR}_{k}>G)$ in \eqref{thoughput_definition}, we first consider the case where the ASE is approximated by another formula associated with the unconditional probability, as for instance in \cite{DBLP:journals/corr/abs-1305-6161} for channel-aware schemes. The approximate ASE in this case is given by
\begin{eqnarray}
\mathcal{T}(\beta)&=&\widetilde{\lambda}\log_{2}(1+\beta) \mathbb{P}(\widetilde{\text{SIR}}_{k}>\beta) \nonumber \\
&=&\lambda P_{s} \frac{e^{-\lambda P_{s} C(\alpha) \beta ^{\frac{2}{\alpha}}}}{1+K(\alpha) \beta^{\frac{2}{\alpha}}}\log_{2}(1+\beta),
\label{thoughput_definition 1}
\end{eqnarray}
where $C(\alpha)= \frac{\pi d_{k,k}^{2} }{\sinc(\frac{2}{\alpha})}$. Although $\mathcal{T}(\beta)$ is not a concave function, its unique maximum point can be found by using the first order optimality condition. Maximizing $\eqref{thoughput_definition 1}$ subject to $0<P_{s}\leq1$ yields the optimal access probability in the unconditional case as follows:
\begin{equation}
\tilde{P_{s}^{\star}} = 
\left\lbrace 
\begin{array}{ccc}
  \frac{1}{\lambda C(\alpha) \beta^{2/\alpha}}
  & \mbox{if} & \lambda >\frac{1}{C(\alpha) \beta^{\frac{2}{\alpha}} }\\
  & & \\
  1
   & \mbox{if} & \lambda \leq \frac{1}{C(\alpha) \beta^{\frac{2}{\alpha}} }.\\
\end{array} \right.
\label{unconditional_Ps}
\end{equation} 

From $\eqref{Ps}$ we have 
\begin{equation}
\exp\left( -\lambda C(\alpha) \tilde{G^{\star}}^{\frac{2}{\alpha}}\right)=\tilde{P_{s}^{\star}} \left(1+K(\alpha) \tilde{G^{\star}}^{\frac{2}{\alpha}}\right),
\label{G_suboptimal}
\end{equation}
where $\tilde{G^{\star}}$ is the optimal SIR threshold that needs to be determined.
For a general type of equation $p^{ax+b}=cx+d$ where $x$ is the variable to be solved and $a$, $b$, $c$, $d$, $p$ are constant, when $p>0$ and $a,c\neq 0$ the solution by using Lambert $W$ function is
\begin{equation}
x=-\frac{W\left(-\frac{a\ln p}{c} p^{b-\frac{ad}{c}}\right)}{a\ln p}-\frac{d}{c}.
\label{lambert}
\end{equation}
From \eqref{unconditional_Ps} we have $\tilde{P_{s}^{\star}} =\frac{1}{\lambda C(\alpha) \beta^{2/\alpha}}$ when $\lambda >\frac{1}{C(\alpha) \beta^{\frac{2}{\alpha}} }$. Putting it into \eqref{G_suboptimal} and using \eqref{lambert} as the typical solution for the type of equation as \eqref{G_suboptimal}, we have the unconditional optimal SIR threshold $\tilde{G^{\star}}$ given by
\begin{equation}
\tilde{G^{\star}}=\left(\frac{W\left(\frac{\lambda^{2} C(\alpha) ^{2} \beta^{\frac{2}{\alpha}} }{K(\alpha)}e^{\frac{\lambda C(\alpha)}{K(\alpha)}}\right)}{\lambda C(\alpha) }-\frac{1}{K(\alpha)}\right)^{\frac{\alpha}{2}},
\label{suboptimal_threshold}
\end{equation}
where $W$ denote Lambert $W$ function. When $\lambda \leq\frac{1}{C(\alpha) \beta^{\frac{2}{\alpha}} }$ the access scheme is not applied thus $\tilde{G^{\star}}= 0$.

In order to see how $\tilde{G^{\star}}$ scales with the target SIR $\beta$, using the expansion of the Lambert function, which is $W(x)\simeq \ln (x)$ when $x\rightarrow \infty$, we have that
\begin{itemize}
\item when $\beta\rightarrow 0$, $\tilde{G^{\star}}\simeq 0$,

\item when $\beta\rightarrow \infty$, 
\begin{equation}
\tilde{G^{\star}}\simeq \left(\frac{\ln \left(\frac{\lambda^{2} C(\alpha)^{2} \beta ^{2/\alpha}}{K(\alpha)}\right)+\frac{\lambda C(\alpha)}{K(\alpha)}}{\lambda C(\alpha)} - \frac{1}{K(\alpha)}\right)^{\frac{\alpha}{2}},
\end{equation}
which means that $\tilde{G^{\star}}\sim \ln(\beta)$.
\end{itemize}

\subsection{SIR-Aware Access Control with Conditional Optimal SIR Threshold}
Given that the SIR distribution of a pre-selected set of active D2D links is conditioned on the SIR threshold $G$ that is used for the access decision, in $\eqref{thoughput_definition}$ the conditional probability $\mathbb{P}(\widetilde{\text{SIR}}_{k}>\beta|\text{SIR}_{k}>G)$ concerns two dependent events, thus is determined by the joint probability of the two events:
\begin{equation}
\mathbb{P}(\widetilde{\text{SIR}}_{k}>\beta|\text{SIR}_{k}>G)
=\frac{\mathbb{P}(\widetilde{\text{SIR}}_{k}>\beta , \text{SIR}_{k}>G )}{P_{s}}.
\label{conditional_probability}
\end{equation}

Although it seems difficult to obtain a neat expression for the joint probability in $\eqref{conditional_probability}$, it is possible to approximate the impact of access probability $P_{s}$ on D2D network throughput in the following two regimes: 
\begin{itemize}
\item if $G\gg\beta$, which implies $P_{s}\rightarrow 0$, the set $\mathcal{A}=\left\{k \in (1, \dots, N) : \text{SIR}_{k}>G \right\}$ can be approximately seen as a subset of $\mathcal{B}=\left\{k \in (1, \dots, N) : \widetilde{\text{SIR}}_{k}>\beta \right\}$, thus\\
\begin{equation}
\mathbb{P}(\widetilde{\text{SIR}}_{k}>\beta , \text{SIR}_{k}>G )\simeq \mathbb{P}(\text{SIR}_{k}>G)=P_{s}
\label{extreme1}
\end{equation}

\item if $G\ll\beta$, which implies $P_{s}\rightarrow 1$, the set $\mathcal{B}=\left\{k \in (1, \dots, N) : \widetilde{\text{SIR}}_{k}>\beta \right\}$ can be approximately seen as a subset of $\mathcal{A}=\left\{k \in (1, \dots, N) : \text{SIR}_{k}>G \right\}$, thus
\begin{eqnarray}
\mathbb{P}(\widetilde{\text{SIR}}_{k}>\beta , \text{SIR}_{k}>G ) &\simeq& \mathbb{P}(\widetilde{\text{SIR}}_{k}>\beta) \nonumber\\
&=& \frac{e^{-\lambda P_{s} C(\alpha)\beta^{\frac{2}{\alpha}}}}{1+K(\alpha) \beta^{\frac{2}{\alpha}}}.
\label{extreme2}
\end{eqnarray}
\end{itemize}

Putting $\eqref{extreme1}$ and $\eqref{extreme2}$ into $\eqref{thoughput_definition}$ and considering $\mathcal{T}(\beta)$ as a function of $P_{s}$ , we have
\begin{equation}
\mathcal{T}(P_s{}) = 
\left\lbrace 
\begin{array}{ccc}
   \lambda P_{s}\log_{2}(1+\beta)
    & \mbox{when} & P_{s}\rightarrow 0\\
  & & \\
   \frac{\lambda\log_{2}(1+\beta)}{1+K(\alpha) \beta^{\frac{2}{\alpha}}} \exp(-\lambda P_{s} C(\alpha)  \beta^{\frac{2}{\alpha}})
    & \mbox{when} &  P_{s}\rightarrow 1. \\
\end{array} \right.
\end{equation} 
Since $\mathcal{T}(P_{s})$ increases monotonically with $P_{s}$ when $P_{s}$ is near $0$, and decreases monotonically with $P_{s}$ when $P_{s}$ is near $1$, and is a continuous function, it is reasonable to consider that the crossing point of these two functions is approximately the $P_{s}$ that maximizes $\mathcal{T}(P_{s})$. Under this assumption, the optimal access probability $P_{s}^{\star}$ verifies
\begin{equation}
P_{s}^{\star} =\exp\left(-\lambda C(\alpha) P_{s}^{\star} \beta^{\frac{2}{\alpha}}\right)\frac{1}{1+K(\alpha) \beta^{\frac{2}{\alpha}}} .
\label{optimal_Probability}
\end{equation}
Similar to the unconditional probability case, by using the Lambert $W$ function, we can obtain an approximate optimal access probability as follows:
\begin{propi}
For a D2D underlaid cellular network with SIR-aware opportunistic access control, the optimal access probability of D2D links which maximizes the D2D network throughput based on the conditional probability approximation is given by
\begin{equation}
P_{s}^{\star}=\min \left\{\frac{W\left(\frac{\lambda C(\alpha) \beta^{\frac{2}{\alpha}}}{1+K(\alpha)\beta^{\frac{2}{\alpha}}}\right)}{\lambda C(\alpha) \beta^{\frac{2}{\alpha}}}, 1\right\}.
\label{proposition}
\end{equation}
\end{propi}
Applying $\eqref{proposition}$ to $\eqref{Ps}$, the conditional optimal SIR threshold is given by
\begin{equation}
G^{\star}=\left(\frac{W\left(\frac{\lambda C(\alpha)}{K(\alpha) P_{s}^{\star}} e^{\frac{\lambda C(\alpha)}{K(\alpha)}}\right)}{\lambda C(\alpha)}-\frac{1}{K(\alpha)}\right)^{\frac{\alpha}{2}}.
\label{optimal_SIR}
\end{equation}

When the SIR-aware opportunistic access scheme is utilized, the optimal SIR thresholds as obtained in \eqref{suboptimal_threshold} and \eqref{optimal_SIR} can be applied directly on each D2D link without centralized control from the BS.

\section{Simulation Results}
In this section, we evaluate the performance of the proposed distributed SIR-aware access control algorithm for D2D underlaid cellular networks, in terms of D2D link coverage probability, area spectral efficiency and average sum rate. The performance of both proposed optimal SIR threshold calculation methods for SIR-based access control is compared to two other access strategies:
\begin{itemize}
\item No access control (No AC): all D2D links are active
\item Channel-aware AC \cite{DBLP:journals/corr/abs-1305-6161}.
\end{itemize}
Simulations are performed in a macrocell of radius $R_c = 500 \text{m}$, where the BS is located at the center of the disk region. The uplink user is uniformly distributed in the coverage region. The locations of D2D transmitters form a spatial PPP with density $\lambda$ according to the value we choose, i.e. $\lambda \in \{2,10\}\times 10^{-5}$. Each D2D receiver is placed randomly around its transmitter at a fixed distance $d_{k,k} = 50\text{m}$. We choose $p_{c}=10 \text{mW}$ and $p_{d}=0.1 \text{mW}$ as transmit powers of uplink user and active D2D transmitter respectively. Pathloss exponent is set at $\alpha=4$. The optimal SIR thresholds are applied as in \eqref{suboptimal_threshold} and \eqref{optimal_SIR}.  All the results are obtained by averaging over $20000$ realizations. 

\subsection{Throughput vs. Link Density $\lambda$ with Target SIR $\beta=5$ dB}
\begin{figure}
\centering
\includegraphics[scale=0.6]{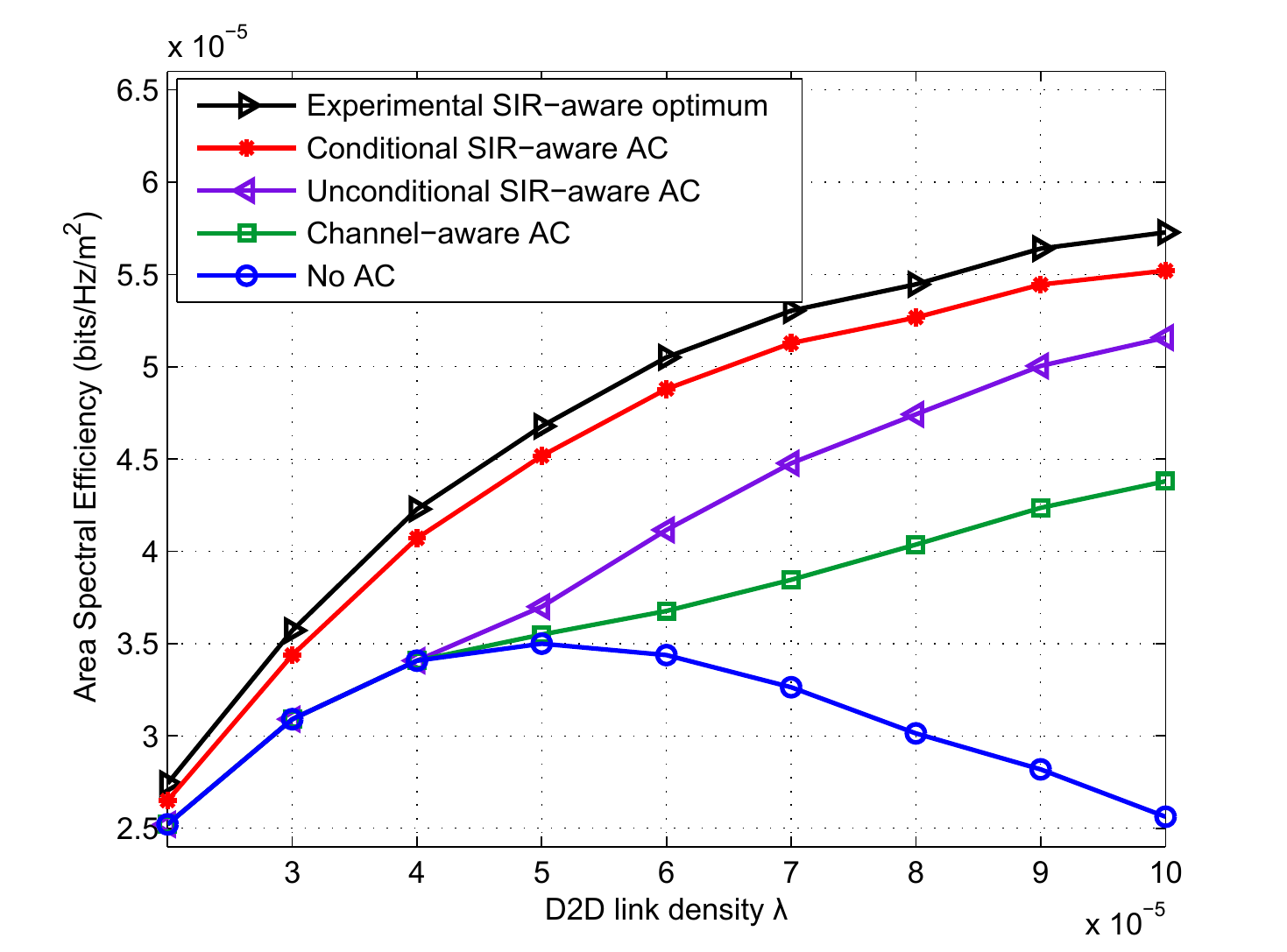}
\caption{Area spectral efficiency of D2D network according to different access control methods. Target SIR $\beta=5\text{dB}$.}
\label{ASE_vs_lambda}
\end{figure}

\begin{figure}
\centering
\includegraphics[scale=0.6]{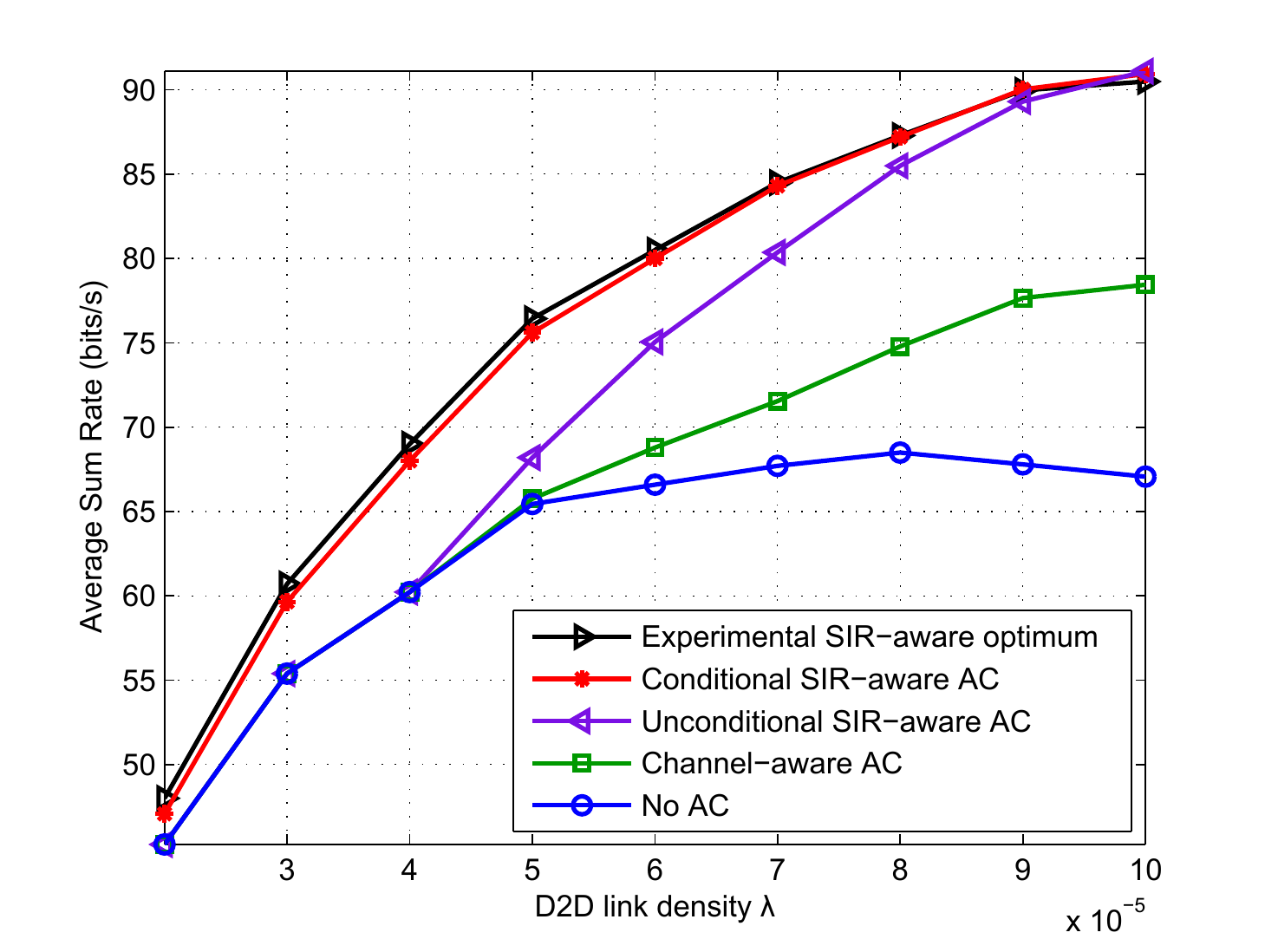}
\caption{Average sum rate of D2D network according to different access control methods. Target SIR $\beta=5\text{dB}$.}
\label{sumrate_vs_lambda}
\end{figure}

Figs. \ref{ASE_vs_lambda} and \ref{sumrate_vs_lambda} show the ASE and average sum rate of the D2D underlaid network as a function of D2D link density $\lambda$ with target SIR $\beta = 5\text{dB}$. We plot also a reference curve: the ASE/average sum rate obtained with experimental optimal SIR threshold by exhaustive searching of the optimal access probability which gives the maximum ASE. We observe that our proposed SIR-aware opportunistic access scheme with both unconditional and conditional optimal SIR thresholds improve the aggregate throughput and provide evident performance gain when compared to channel-aware AC and, obviously, to the case with no AC. As expected, the conditional optimal SIR threshold scheduling has better performance than the unconditional methodology, since the former exploits the dependence between the SIR distribution of selected active D2D links and that of all potential D2D links, i.e. it captures more accurately the resulting performance of threshold scheduling than in the unconditional case. When compared to the experimental optimum of SIR-aware AC, the conditional optimal SIR threshold achieves very close performance.

We also observe that the SIR-aware strategy with conditional optimal SIR threshold improves the network performance for any range of D2D densities and is always superior to all other access schemes, while the two alternative methods start having gains from $\lambda=4\times 10^{-5}$. This result implies that SIR-aware access control with conditional SIR threshold exploits better potential throughput increase by taking into account the link quality condition of active D2D links.

\subsection{Throughput vs. Target SIR in Sparse and Dense D2D Networks}
\begin{figure}
\centering
\includegraphics[scale=0.55]{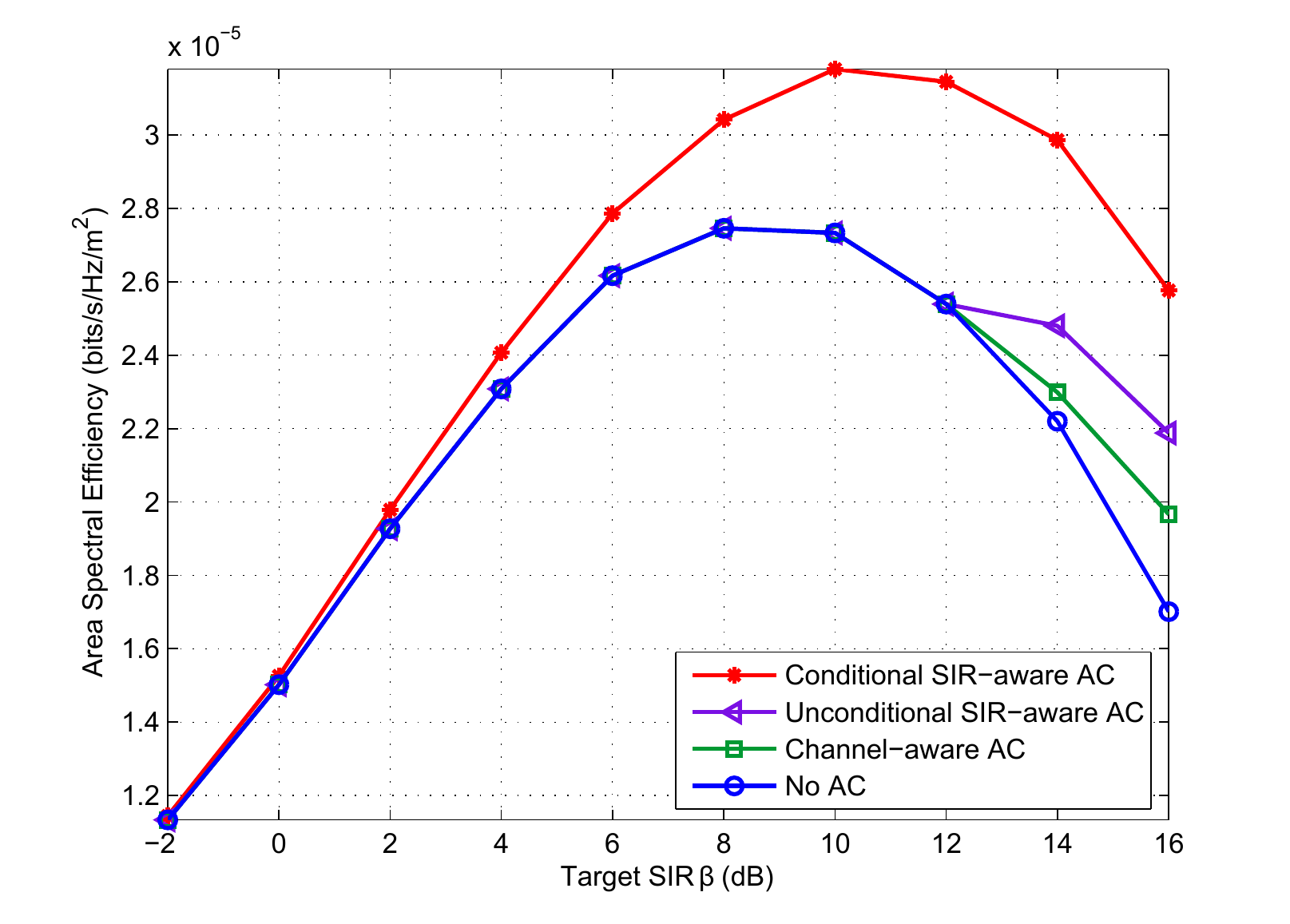}
\caption{Area spectral efficiency of dense D2D network according to different access control methods, i.e., $\lambda=2\times 10^{-5}$.}
\label{ASE_vs_beta_sparse}
\end{figure}

\begin{figure}
\centering
\includegraphics[scale=0.55]{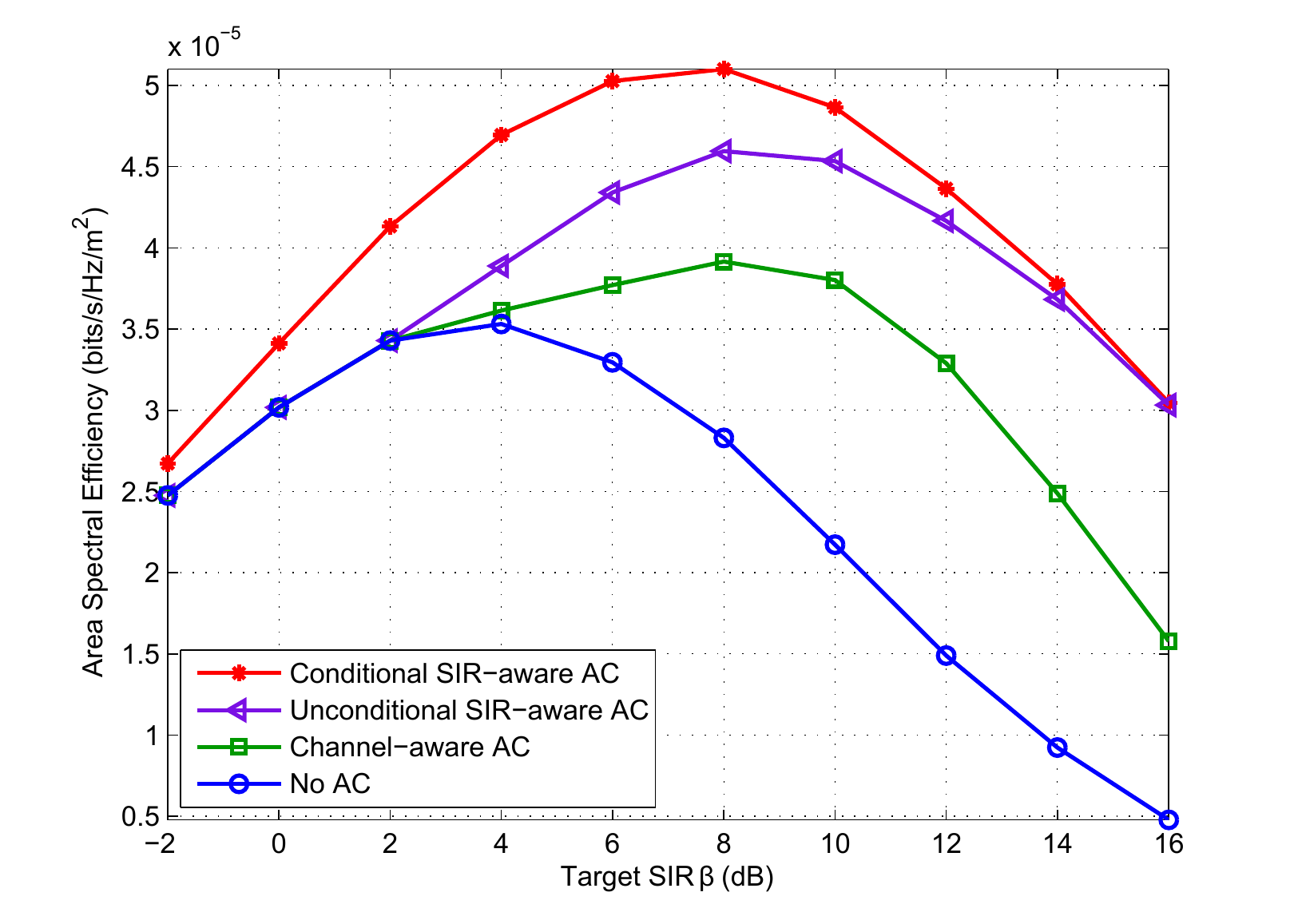}
\caption{Area spectral efficiency of dense D2D network according to different access control methods, i.e., $\lambda=6\times 10^{-5}$.}
\label{ASE_vs_beta_dense}
\end{figure}

Figs. \ref{ASE_vs_beta_sparse} and \ref{ASE_vs_beta_dense} show the ASE of D2D network as a function of target SIR $\beta$ in both sparse (i.e. $\lambda=2\times 10^{-5}$) and dense (i.e. $\lambda=6\times 10^{-5}$) D2D network scenarios. Channel-aware and SIR-aware AC with unconditional optimal SIR threshold are applied when the target SIR is beyond $\tilde{\beta^{\star}}=\left(\lambda C(\alpha)\right)^{-\frac{\alpha}{2}}$, that is, $12$ dB for the sparse D2D scenario and $2$ dB for the dense one. However, the SIR-aware AC with conditional optimal threshold is activated even before $\tilde{\beta^{\star}} = -2$ dB. Furthermore, decentralized SIR-aware access scheme with conditional SIR threshold exhibits performance gains in both sparse and dense D2D network configurations, while in dense D2D networks, the performance gap between SIR-aware AC with conditional and unconditional SIR threshold is reduced for $\beta > 12$ dB. This implies that the approximation error caused by ignoring the link quality condition of selected D2D links becomes negligible in high link density and high target SIR regimes, thus the unconditional optimal access probability $\tilde{P_{s}^{\star}}=\min\left\{\frac{\sinc(2/\alpha)}{\pi \lambda \beta^{\frac{2}{\alpha}} d_{k,k}^{2}},1\right\}$ becomes near optimal.

\subsection{Coverage Probability in Sparse and Dense D2D networks}
\begin{figure}
\centering
\includegraphics[scale=0.55]{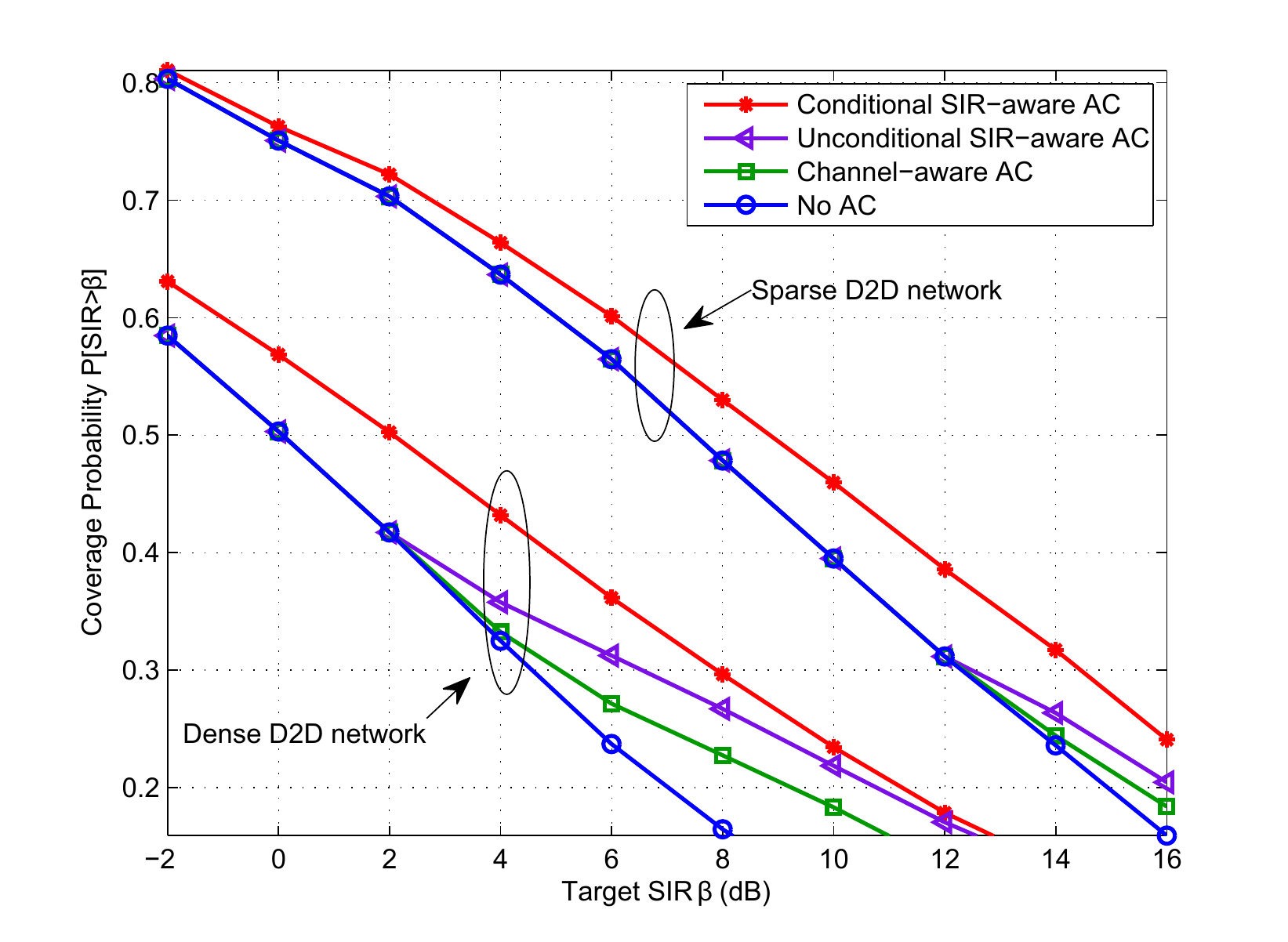}
\caption{Coverage probability of D2D links according to different access control methods. For the sparse D2D scenario, $\lambda=2\times 10^{-5}$, for the dense one, $\lambda=6\times 10^{-5}$ .}
\label{Pcov_vs_beta}
\end{figure}

In Fig. \ref{Pcov_vs_beta}, we plot the D2D link coverage probability as a function of target SIR $\beta$ in sparse and dense D2D network settings. Similar to the previous results, we observe the benefits of using SIR-aware opportunistic AC, especially in the high target SIR regime and in dense D2D underlaid cellular networks. Moreover, as far as the optimal SIR threshold calculation methodology is concerned, use of the conditional SIR threshold results in higher performance gain, which also verifies our theoretical analysis.

\section{Conclusions}
In this work, we proposed a decentralized SIR-aware opportunistic access control algorithm for D2D underlaid cellular networks with optimal activation probability calculated to maximize the area spectral efficiency of D2D links. We provide two approximate expressions for the optimal SIR threshold based on the unconditional and conditional formulation of D2D success probability. The performance gain achieved by our SIR-aware access control algorithm is evaluated in terms of D2D link coverage probability and D2D network throughput. The main takeaway of this paper is that SIR-aware opportunistic access with adaptive SIR threshold obtained by throughput optimization based on the conditional success probability provides the best system performance improvement compared to alternative spatial opportunistic access schemes.




\end{document}